\documentclass[10pt]{article}
\usepackage{graphicx}
\usepackage{amsmath}
\usepackage{amssymb}
\usepackage{caption2}
\begin{document}
\begin{center}
{\large {\bf \sc{  Vector hidden-bottom tetraquark candidate: $Y(10750)$   }}} \\[2mm]
Zhi-Gang  Wang \footnote{E-mail: zgwang@aliyun.com.  }     \\
 Department of Physics, North China Electric Power University, Baoding 071003, P. R. China
\end{center}

\begin{abstract}
In this article, we take the scalar diquark and antidiquark operators as the basic constituents, construct  the $C\gamma_5\otimes\stackrel{\leftrightarrow}{\partial}_\mu\otimes \gamma_5C$ type tetraquark current  to study  the $Y(10750)$  with the QCD sum rules.  The  predicted mass $M_{Y}=10.75\pm0.10\,\rm{GeV}$ and width $\Gamma_Y= 33.60^{+16.64}_{-9.45}\,{\rm{MeV}}$ support assigning the $Y(10750)$  to be the  diquark-antidiquark type vector hidden-bottom tetraquark state, which has  a relative P-wave between the diquark and antidiquark constituents.
 \end{abstract}



\section{Introduction}

Recently, the Belle collaboration observed a resonance structure $Y(10750)$ with the global significance of $6.7 \sigma$ in the  $e^+e^-\to\Upsilon(nS)\pi^+\pi^-
$ ($n=1,\,2,\,3$) cross sections at energies from $10.52$ to $11.02\,\rm{ GeV}$ using the data collected with the Belle detector at the KEKB asymmetric-energy
 $e^+e^-$  collider \cite{Belle-Y10750}. The Breit-Wigner  mass and width are
$M_Y=10752.7\pm5.9\,{}^{+0.7}_{-1.1}\,\rm{MeV}$ and $\Gamma_Y=35.5^{+17.6}_{-11.3}\,{}^{+3.9}_{-3.3}\,\rm{MeV}$, respectively. The $Y(10750)$ is observed in the
 processes $Y(10750)\to\Upsilon(nS)\pi^+\pi^-
$ ($n=1,\,2,\,3$), its quantum numbers may be $J^{PC}=1^{--}$.   In the famous Godfrey-Isgur model, the nearby bottomonium states are the $\Upsilon({\rm 4S})$, $\Upsilon({\rm 5S})$ and $\Upsilon({\rm 3D})$ with the masses $10.635\,\rm{GeV}$, $10.878\,\rm{GeV}$ and $10.698\,\rm{GeV}$, respectively \cite{Godfrey-2015}, while
in the QCD-motivated relativistic quark model based on the quasipotential approach (the screened potential model), the corresponding masses are $10.586\,\rm{GeV}$, $10.869\,\rm{GeV}$ and $10.704\,\rm{GeV}$ ($10.611\,\rm{GeV}$, $10.831\,\rm{GeV}$ and $10.670\,\rm{GeV}$ \cite{ChaoKT}), respectively \cite{Ebert-2011}. Without introducing mixing effects, the experimental data $M_Y=10752.7\pm5.9\,{}^{+0.7}_{-1.1}\,\rm{MeV}$ cannot be reproduced, if we assign the $Y(10750)$ to be a conventional  bottomonium state \cite{Zhong-Y10750}.

The $Y(10750)$ may be a hidden-bottom tetraquark candidate.  In Refs.\cite{Wang-Y4260-D,Wang-Vector-D},  we take the scalar  and axialvector diquark operators as the basic constituents, as they are favored quark configurations,  introduce a relative  P-wave between the scalar (or axialvector) diquark and scalar (or axialvector) antidiquark operators explicitly in constructing the vector tetraquark current operators,
 and calculate  the masses and pole residues of the  vector hidden-charm tetraquark states using  the QCD sum rules in a systematic way, and obtain the lowest masses of the vector hidden-charm tetraquark states  up to now,
 the predictions  support  assigning the exotic states  $Y(4220/4260)$,  $Y(4320/4360)$, $Y(4390)$ and $Z(4250)$ to be the vector tetraquarks with the quantum numbers $J^{PC}=1^{--}$, which originate from the  relative P-wave between the diquark and antidiquark constituents. On the other hand, if we take the scalar ($C\gamma_5$-type), pseudoscalar ($C$-type), vector ($C\gamma_\alpha\gamma_5$-type) and axialvector ($C\gamma_\alpha$-type) diquark operators as the basic constituents,  and construct the vector tetraquark current operators having the quantum numbers $J^{PC}=1^{--}$ without introducing the relative P-wave between the diquark and antidiquark constituents, we can obtain the masses of the lowest vector tetraquark states, which are about $4.34\,\rm{GeV}$ or $4.59\,\rm{GeV}$  \cite{Wang-Vector-EPJC},  and are larger or much larger than the measured  mass of the $Y(4220/4260)$ from the BESIII collaboration \cite{BES-Y4220-Y4320},  because the pseudoscalar and vector diquarks are not favored quark  configurations  \cite{Wang-Vector-EPJC}.
 In Ref.\cite{Wang-Hidden-Bottom},  we take the scalar and axialvector diquark (and antidiquark) operators as the basic constituents to construct the current operators, calculate the  masses and pole residues  of the hidden-bottom tetraquark states  with  the quantum numbers $J^{PC}=0^{++}$, $1^{++}$, $1^{+-}$ and $2^{++}$ systematically using   the QCD sum rules, and observe that the masses of the ground state hidden-bottom tetraquark states  are about $10.61-10.65\,\rm{GeV}$. The $Y(10750)$ may be a vector hidden-bottom tetraquark state.

In the present work, we  tentatively assign the $Y(10750)$ as a diquark-antidiquark type vector hidden-bottom tetraquark state with the quantum numbers  $J^{PC}=1^{--}$, and construct  the $C\gamma_5\otimes\stackrel{\leftrightarrow}{\partial}_\mu\otimes \gamma_5C$ type tetraquark current operator to calculate its mass and pole residue using the QCD sum rules. In calculations, we take into account the vacuum condensates up to dimension 10  in  the operator product expansion as in our previous works.
 Furthermore, we study the  two-body strong decays of the vector hidden-bottom tetraquark candidate $Y(10750)$ with the three-point correlation functions by carrying out the operator product expansion up to the vacuum condensates of dimension $5$. In calculations, we take into account both the connected and disconnected Feynman diagrams.

The rest of the paper is organized as follows. In section 2, we obtain the QCD sum rules for  the mass and pole residue  of  the $Y(10750)$;
In section 3,  we obtain the QCD sum rules for  the hadronic coupling constants in the strong decays of  the $Y(10750)$, then obtain the partial decay widths; Section 4 is given for a short conclusion.

\section{The  mass and pole residue of the  vector tetraquark candidate $Y(10750)$}
Firstly, we write down  the two-point correlation function $\Pi_{\mu\nu}(p)$  in the QCD sum rules,
\begin{eqnarray}
\Pi_{\mu\nu}(p)&=&i\int d^4x e^{ip \cdot x} \langle0|T\left\{J_\mu(x)J_\nu^{\dagger}(0)\right\}|0\rangle \, ,
\end{eqnarray}
where $J_\mu(x)=J_\mu^{(1,\pm1)}(x)$, $J_\mu^{(1,0)}(x)$ and $J_\mu^{(0,0)}(x)$,
\begin{eqnarray}
J^{(1,1)}_\mu(x)&=&\frac{\varepsilon^{ijk}\varepsilon^{imn}}{\sqrt{2}}u^{Tj}(x)C\gamma_5 b^k(x)\stackrel{\leftrightarrow}{\partial}_\mu \bar{d}^m(x)\gamma_5 C \bar{b}^{Tn}(x) \, ,\nonumber \\
J^{(1,0)}_\mu(x)&=&\frac{\varepsilon^{ijk}\varepsilon^{imn}}{2}\Big[u^{Tj}(x)C\gamma_5 b^k(x)\stackrel{\leftrightarrow}{\partial}_\mu \bar{u}^m(x)\gamma_5 C \bar{b}^{Tn}(x)\nonumber\\
&&- \,d^{Tj}(x)C\gamma_5 b^k(x)\stackrel{\leftrightarrow}{\partial}_\mu \bar{d}^m(x)\gamma_5 C \bar{b}^{Tn}(x) \Big] \, , \nonumber\\
J^{(1,-1)}_\mu(x)&=&\frac{\varepsilon^{ijk}\varepsilon^{imn}}{\sqrt{2}}d^{Tj}(x)C\gamma_5 b^k(x)\stackrel{\leftrightarrow}{\partial}_\mu \bar{u}^m(x)\gamma_5 C \bar{b}^{Tn}(x) \, , \nonumber\\
J^{(0,0)}_\mu(x)&=&\frac{\varepsilon^{ijk}\varepsilon^{imn}}{2}\Big[u^{Tj}(x)C\gamma_5 b^k(x)\stackrel{\leftrightarrow}{\partial}_\mu \bar{u}^m(x)\gamma_5 C \bar{b}^{Tn}(x)\nonumber\\
&&+ \,d^{Tj}(x)C\gamma_5 b^k(x)\stackrel{\leftrightarrow}{\partial}_\mu \bar{d}^m(x)\gamma_5 C \bar{b}^{Tn}(x) \Big] \, ,
\end{eqnarray}
where the $i$, $j$, $k$, $m$, $n$ are color indexes, the superscripts $(1,\pm1)$, $(1,0)$, $(0,0)$ denote the isospin indexes $(I,I_3)$, $\stackrel{\leftrightarrow}{\partial}_\mu=\overrightarrow{\partial}_\mu-\overleftarrow{\partial}_\mu$.
In the isospin limit, i.e. $m_u=m_d$, the current operators $J_\mu(x)$ couple potentially  to the  diquark-antidiquark type vector hidden-bottom tetraquark states which have degenerate masses. In the present work,  we choose $J_\mu(x)=J^{(1,1)}_\mu(x)$ for simplicity.

The scattering amplitude for one-gluon exchange  is proportional to
\begin{eqnarray}
\left(\frac{\lambda^a}{2}\right)_{ij}\left(\frac{\lambda^a}{2}\right)_{kl}&=&-\frac{1}{3}\left(\delta_{ij}\delta_{kl}-\delta_{il}\delta_{kj}\right)
+\frac{1}{6}\left(\delta_{ij}\delta_{kl}+\delta_{il}\delta_{kj}\right) \, ,
\end{eqnarray}
where
\begin{eqnarray}
\varepsilon^{mik}\varepsilon^{mjl} &=&\delta_{ij}\delta_{kl}-\delta_{il}\delta_{kj}\, ,
\end{eqnarray}
the $\lambda^a$ is the  Gell-Mann matrix.  The negative (positive) sign in front of the antisymmetric  antitriplet $\bar{3}_c$ (symmetric sextet $6_c$) indicates the interaction is attractive (repulsive), which favors (disfavors)  formation of
the diquarks in  color antitriplet $\bar{3}_c$ (color sextet $6_c$). We prefer the  diquark operators in  color antitriplet $\bar{3}_c$ to the diquark operators in  color sextet  $6_c$ in constructing the tetraquark current operators to interpolate the lowest tetraquark states.

At the phenomenological  side, we  take into account the non-vanishing current-hadron  couplings considering the same quantum numbers, and  separate  the
contribution of the ground state vector hidden-bottom tetraquark state in correlation function $\Pi_{\mu\nu}(p)$ \cite{SVZ79,Reinders85}, which is supposed to be  the $Y(10750)$,
\begin{eqnarray}
\Pi_{\mu\nu}(p)&=&\frac{\lambda_{Y}^2}{M_{Y}^2-p^2}\left(-g_{\mu\nu} +\frac{p_\mu p_\nu}{p^2}\right)  +\cdots \, ,
\end{eqnarray}
where the pole residue  $\lambda_{Y}$ is  defined by $\langle 0|J_\mu(0)|Y(p)\rangle=\lambda_{Y} \,\varepsilon_\mu$,
the $\varepsilon_\mu$ is the polarization vector.

At the QCD side, we carry out the operator product expansion up to the vacuum condensates of dimension 10 in a consistent way, and take into account the vacuum condensates $\langle\bar{q}q\rangle$, $\langle\frac{\alpha_{s}GG}{\pi}\rangle$, $\langle\bar{q}g_{s}\sigma Gq\rangle$, $\langle\bar{q}q\rangle^2$,
$\langle\bar{q}q\rangle \langle\frac{\alpha_{s}GG}{\pi}\rangle$,  $\langle\bar{q}q\rangle  \langle\bar{q}g_{s}\sigma Gq\rangle$,
$\langle\bar{q}g_{s}\sigma Gq\rangle^2$ and $\langle\bar{q}q\rangle^2 \langle\frac{\alpha_{s}GG}{\pi}\rangle$, then obtain the QCD spectral density through dispersion relation,  take the
quark-hadron duality below the continuum threshold  $s_0$ and perform the Borel transform   to obtain  the  QCD sum rules:
\begin{eqnarray}\label{QCDSR}
\lambda^2_{Y}\, \exp\left(-\frac{M^2_{Y}}{T^2}\right)= \int_{4m_b^2}^{s_0} ds\, \rho(s) \, \exp\left(-\frac{s}{T^2}\right) \, .
\end{eqnarray}
For the explicit expression of the QCD spectral density $\rho(s)$ and the technical details in calculating the Feynman diagrams, one can consult Refs.\cite{Wang-Y4260-D,WangHuangtao-2014-PRD}.

Then we can obtain the QCD sum rules for the mass of the vector hidden-bottom tetraquark candidate $Y(10750)$ through a fraction,
 \begin{eqnarray}\label{QCDSR-mass}
 M^2_{Y}&=& -\frac{\int_{4m_b^2}^{s_0} ds\frac{d}{d \tau}\,\rho(s)\,\exp\left(-\tau s \right)}{\int_{4m_b^2}^{s_0} ds \,\rho(s)\,\exp\left(-\tau s\right)}\mid_{\tau=\frac{1}{T^2}}\, .
\end{eqnarray}

We choose  the conventional  values (in other words,  the popular values) of the vacuum condensates $\langle
\bar{q}q \rangle=-(0.24\pm 0.01\, \rm{GeV})^3$,   $\langle
\bar{q}g_s\sigma G q \rangle=m_0^2\langle \bar{q}q \rangle$,
$m_0^2=(0.8 \pm 0.1)\,\rm{GeV}^2$,  $\langle \frac{\alpha_s
GG}{\pi}\rangle=(0.33\,\rm{GeV})^4 $    at the energy scale  $\mu=1\, \rm{GeV}$
\cite{SVZ79,Reinders85,Colangelo-Review}, and take the $\overline{MS}$ mass $m_{b}(m_b)=(4.18\pm0.03)\,\rm{GeV}$ listed in "The Review of Particle Physics"  \cite{PDG}, and set the $u$ and $d$ quark masses to be zero.
Furthermore, we take into account  the energy-scale dependence of  the  parameters  at the QCD side from the renormalization group equation  \cite{Narison-mix},
\begin{eqnarray}
\langle\bar{q}q \rangle(\mu)&=&\langle\bar{q}q \rangle({\rm 1GeV})\left[\frac{\alpha_{s}({\rm 1GeV})}{\alpha_{s}(\mu)}\right]^{\frac{12}{33-2n_f}}\, , \nonumber\\
 \langle\bar{q}g_s \sigma Gq \rangle(\mu)&=&\langle\bar{q}g_s \sigma Gq \rangle({\rm 1GeV})\left[\frac{\alpha_{s}({\rm 1GeV})}{\alpha_{s}(\mu)}\right]^{\frac{2}{33-2n_f}}\, , \nonumber\\
 m_b(\mu)&=&m_b(m_b)\left[\frac{\alpha_{s}(\mu)}{\alpha_{s}(m_b)}\right]^{\frac{12}{33-2n_f}} \, ,\nonumber\\
\alpha_s(\mu)&=&\frac{1}{b_0t}\left[1-\frac{b_1}{b_0^2}\frac{\log t}{t} +\frac{b_1^2(\log^2{t}-\log{t}-1)+b_0b_2}{b_0^4t^2}\right]\, ,
\end{eqnarray}
   where $t=\log \frac{\mu^2}{\Lambda^2}$, $b_0=\frac{33-2n_f}{12\pi}$, $b_1=\frac{153-19n_f}{24\pi^2}$, $b_2=\frac{2857-\frac{5033}{9}n_f+\frac{325}{27}n_f^2}{128\pi^3}$,  $\Lambda=210\,\rm{MeV}$, $292\,\rm{MeV}$  and  $332\,\rm{MeV}$ for the flavors  $n_f=5$, $4$ and $3$, respectively \cite{PDG}. In the present work, as we study the vector hidden-bottom tetraquark state, it is better to choose the flavor $n_f=5$,  then evolve all the input parameters  to the ideal  energy scale $\mu$.

The Borel parameter $T^2$ is a free parameter, the continuum threshold parameter $s_0$ is also a free parameter, but we
can borrow some ideas from  the mass spectrum of the conventional mesons and the established exotic mesons and put additional constraints on the $s_0$
so as to avoid contaminations from the excited states and continuum states. In the conventional QCD sum rules, there are two  basic criteria (i.e. "pole dominance at the hadron side" and "convergence of the operator product expansion") to obey. In the QCD sum rules for the multiquark states, we add  two additional criteria,
 (i.e. "appearance of the flat Borel platforms" and "satisfying the modified energy scale formula"), as in the QCD sum rules for the conventional mesons and baryons, we cannot obtain very flat Borel platforms  due to lacking  higher dimensional vacuum  condensates to stabilize  the QCD sum rules. Now we search for the optimal values of the  two parameters  to satisfy the  four  criteria  via try and error.

 In Refs.\cite{WangHuangtao-2014-PRD,Wang-tetra-formula,Wang-tetra-NPA}, we study   the hidden-charm and hidden-bottom tetraquark states (which  consist of a diquark-antidiquark pair in relative S-wave) with the QCD sum rules, and explore the energy scale dependence of the extracted masses and pole residues for the first time.

 In the heavy quark limit $m_Q \to \infty$, the heavy  quark $Q$ serves as a static well potential and  attracts  the light quark $q$ to form a diquark in the color antitriplet $\bar{3}_c$, while  the heavy  antiquark $\overline{Q}$ serves as
another static well potential and attracts  the light antiquark $\bar{q}$ to form a antidiquark in the color triplet $3_c$. Then the diquark and antidiquark attract  each other to form a compact tetraquark state.

The favored heavy diquark configurations are the scalar and axialvector diquark operators $\varepsilon^{ijk}q^{Tj}C\gamma_5Q^k$ and $\varepsilon^{ijk}q^{Tj}C\gamma_\alpha Q^k$ in the color antitriplet $\bar{3}_c$ \cite{WangDiquark}.    If there exists an additional P-wave between the light quark and heavy quark, we can obtain the pseudoscalar and vector diquark operators $\varepsilon^{ijk}q^{Tj}C\gamma_5 \underline{\gamma_5}Q^k$ and $\varepsilon^{ijk}q^{Tj}C\gamma_\alpha\underline{\gamma_5} Q^k$ in the color antitriplet without introducing the additional P-wave explicitly, as multiplying a $\gamma_5$ can change the parity, the P-wave effect is embodied in the underlined $\gamma_5$. On the other hand, we can introduce the P-wave  explicitly, and obtain the vector and tensor  diquark operators $\varepsilon^{ijk}q^{Tj}C\gamma_5\stackrel{\leftrightarrow}{\partial}_\alpha Q^k$ and $\varepsilon^{ijk}q^{Tj}C\gamma_\alpha \stackrel{\leftrightarrow}{\partial}_\beta Q^k$ in the color antitriplet.

We can take the $C$, $C\gamma_5$, $C\gamma_\alpha$, $C\gamma_\alpha\gamma_5$, $C\gamma_5\stackrel{\leftrightarrow}{\partial}_\alpha$ and
$C\gamma_\alpha\stackrel{\leftrightarrow}{\partial}_\beta$ type diquark and  antidiquark operators (also the $C\sigma_{\alpha\beta}$ and $C\sigma_{\alpha\beta}\gamma_5$ type diquark operators, which have both $J^{P}=1^+$ and $1^-$ components) as the basic constituents to construct the tetraquark current operators with the $J^{PC}=0^{++}$, $1^{++}$, $1^{+-}$, $1^{--}$ and  $2^{++}$ to interpolate the hidden-charm or hidden-bottom tetraquark states, the P-wave lies
between  the light quark and heavy quark (or between the light antiquark and heavy antiquark) if any, in other words, the P-wave lies inside the diquark or antidiquark, while the diquark and antidiquark are in relative S-wave \cite{Wang-Vector-EPJC,Wang-Hidden-Bottom,WangHuangtao-2014-PRD,Wang-tetra-formula,Wang-tetra-NPA}.
  In this case, we  introduce the effective heavy quark mass  ${\mathbb{M}}_Q$ and virtuality $V=\sqrt{M^2_{X/Y/Z}-(2{\mathbb{M}}_Q)^2}$ to characterize the tetraquark states, and suggest an energy scale formula $\mu=V=\sqrt{M^2_{X/Y/Z}-(2{\mathbb{M}}_Q)^2}$  to choose the optimal  energy scales of the QCD spectral densities \cite{WangHuangtao-2014-PRD,Wang-tetra-formula,Wang-tetra-NPA}.

 On the other hand, if there exists   a relative  P-wave, which lies  between the diquark and antidiquark constituents, we have to consider  the effect of the  P-wave and modify the energy scale formula,
\begin{eqnarray}
\mu&=&\sqrt{M^2_{X/Y/Z}-(2{\mathbb{M}}_Q+{\rm P_E})^2}\, .
\end{eqnarray}
where the $\rm{P_E}$ denotes the energy costed  by the relative P-wave \cite{Wang-Y4260-D,Wang-Vector-D}.
 The $Y(10750)$ lies near the $\Upsilon(\rm 4S)$ and $\Upsilon(\rm 5S)$, the energy gap between the masses of the $\chi_{b1}({\rm 4P})$ and $\Upsilon({\rm 4S})$ ($\chi_{b1}({\rm 5P})$ and $\Upsilon({\rm 5S})$) is about $0.14-0.15\,\rm{GeV}$ ($0.12-0.14\,\rm{GeV}$) in the potential models \cite{Godfrey-2015,ChaoKT}.
 In this article, we study the vector hidden-bottom tetraquark  state, there exists a relative P-wave between the bottom diquark and bottom antidiquark constituents. In the present case,  the relative P-wave is estimated to cost  about $0.12\,\rm{GeV}$, we can modify the energy scale formula to be,
\begin{eqnarray}
\mu&=&\sqrt{M^2_{Y}-(2{\mathbb{M}}_b+0.12\,\rm{GeV})^2}=\sqrt{M^2_{Y}-(10.46\,\rm{GeV})^2}\, ,
\end{eqnarray}
where we choose  the updated value ${\mathbb{M}}_b=5.17\,\rm{GeV}$ \cite{WangZG-X5568}. The value $\rm{P_E}=0.12\,\rm{GeV}$ is reasonable, as the QCD sum rules indicate that the ground state hidden-bottom tetraquark mass is  about  $10.61-10.65\,\rm{GeV}$ \cite{Wang-Hidden-Bottom}, the vector hidden-bottom tetraquark mass is estimated to be $10.73-10.77\,\rm{GeV}$, which is in excellent agreement with (at least is compatible with) the experimental data  $M_Y=10752.7\pm5.9\,{}^{+0.7}_{-1.1}\,\rm{MeV}$ from the Belle    collaboration \cite{Belle-Y10750}.

In Ref.\cite{Wang-Hidden-Bottom}, we study the scalar, axialvector and tensor diquark-antidiquark type hidden-bottom tetraquark states $Z_b$ (where the bottom diquark and bottom antidiquark are in relative S-wave) with the QCD sum rules  systematically, and choose the continuum threshold  parameters as $\sqrt{s_0}=M_{Z_b}+0.55\pm0.10\,\rm{GeV}$, which works well and is consistent with the assumption $M_{Z^\prime_b}-M_{Z_b}=M_{\Upsilon^\prime}-M_{\Upsilon}=0.55\,\rm{GeV}$ \cite{PDG}.   In this article, we assume $M_{Y^\prime}-M_{Y}=M_{\Upsilon^\prime}-M_{\Upsilon}=0.55\,\rm{GeV}$ and choose the continuum threshold parameter as  $\sqrt{s_0}=M_Y+0.55\pm0.10\,\rm{GeV}$.

 In numerical calculations, we observe that the continuum threshold parameter $\sqrt{s_0}=11.3\pm 0.1\,\rm{GeV}$, Borel parameter $T^2=(6.3-7.3)\,\rm{GeV}^2$ and energy scale $\mu=2.5\,\rm{GeV}$ work well. The pole contribution from the ground state tetraquark candidate  $Y(10750)$ is about $(47-70)\%$, the pole dominance is well satisfied. The predicted mass is about $M_{Y}=10.75\,\rm{GeV}$, which certainly obeys the modified energy scale formula.

In numerical calculations, we observe that the contributions of the vacuum condensates $\langle\bar{q}q\rangle$, $\langle\bar{q}g_{s}\sigma Gq\rangle$, $\langle\bar{q}q\rangle^2$ and   $\langle\bar{q}q\rangle  \langle\bar{q}g_{s}\sigma Gq\rangle$ are large, the values  change quickly with variation of the Borel parameter $T^2$ in the region $T^2<6.3\,\rm{GeV}^2$, the convergent behavior is  bad,   we have to choose     $T^2\geq 6.3\,\rm{GeV^2}$. At the Borel window, $T^2=(6.3-7.3)\,\rm{GeV}^2$,  the contributions of the vacuum condensates $\langle\bar{q}q\rangle$, $\langle\bar{q}g_{s}\sigma Gq\rangle$, $\langle\bar{q}q\rangle^2$ and   $\langle\bar{q}q\rangle  \langle\bar{q}g_{s}\sigma Gq\rangle$ have the hierarchy $D_3\gg |D_5|\sim D_6\gg |D_8|$, where we use the symbol   $D_n$ to denote  the contributions of the vacuum condensates of dimension $n$. The contributions of the vacuum condensates  $\langle\frac{\alpha_{s}GG}{\pi}\rangle$ and
$\langle\bar{q}q\rangle \langle\frac{\alpha_{s}GG}{\pi}\rangle$ are very small and cannot affect the convergent behavior of the operator product expansion, the contribution of the vacuum condensates of dimension 10 is    $(2-6)\%$. We can obtain the conclusion that   the operator product expansion is well convergent.

Now we obtain the numerical values of the mass and pole residue of
 the tetraquark candidate    $Y(10750)$ from the QCD sum rules in Eqs.\eqref{QCDSR}-\eqref{QCDSR-mass},  and take into account all uncertainties of the input parameters,
 and plot the predicted mass and pole residue   with variations of the Borel parameter $T^2$ explicitly in Figs.1-2,
\begin{eqnarray}
M_{Y}&=&10.75\pm0.10\,\rm{GeV} \, ,  \nonumber\\
\lambda_{Y}&=&\left( 1.89 \pm0.31 \right) \times 10^{-1}\,\rm{GeV}^6 \,   .
\end{eqnarray}
 It is obvious that there appear platforms in the lines of  both the mass and pole residue in the Borel window, see Figs.1-2. Now the four criteria of the QCD sum rules for the vector tetraquark states are all satisfied \cite{WangHuangtao-2014-PRD,Wang-tetra-formula,Wang-tetra-NPA}, and we expect to make reliable or reasonable predictions.

 The numerical value  $M_{Y}=10.75\pm0.10\,\rm{GeV}$ from the present QCD sum rules is in excellent agreement with (at least is compatible with) the experimental data   $M_Y=10752.7\pm5.9\,{}^{+0.7}_{-1.1}\,\rm{MeV}$ from the Belle    collaboration \cite{Belle-Y10750} (see Fig.1), which favors  assigning the $Y(10750)$  as the diquark-antidiquark type vector hidden-bottom tetraquark state, which has  a relative P-wave between the diquark and antidiquark constituents. The  relative P-wave between the diquark and antidiquark constituents  hampers the rearrangements of the quarks and antiquarks in the color and Dirac-spinor spaces to form the quark-antiquark type meson pairs, which   can interpret  (is compatible with) the small experimental value of the width $\Gamma_Y=35.5^{+17.6}_{-11.3}\,{}^{+3.9}_{-3.3}\,\rm{MeV}$ \cite{Belle-Y10750}.

At the charm sector, the calculations based on the QCD sum rules favors  assigning the
 $Y(4220/4260)$,  $Y(4320/4360)$ and $Y(4390)$  to be the vector tetraquark   states with a relative P-wave between the scalar (or axialvector) diquark and scalar (or axialvector) antidiquark pair \cite{Wang-Y4260-D,Wang-Vector-D}. Furthermore, the  QCD sum rules  favors assigning the $X^*(3860)$ to be the  scalar-diquark-scalar-antidiquark type scalar tetraquark state, where the diquark and antidiquark constituents are in relative S-wave \cite{WangZG-X3860}.
  Analogous arguments survive  both in the bottom and charm sectors, however, unambiguous assignments  call for more experimental data and more theoretical works.

In Fig.\ref{mass-mu}, we plot the predicted mass  of the vector hidden-bottom tetraquark candidate $Y(10750)$  with variation of the energy scale $\mu$ for central values of the input parameters. From the figure, we can see that the predicted mass decreases monotonically and quickly with the increase of the energy scale $\mu$. If we
abandon the modified energy scale formula $\mu=\sqrt{M^2_{X/Y/Z}-(2{\mathbb{M}}_Q+{\rm P_E})^2}$, we are puzzled about which energy scale should be chosen. If we choose  the  typical energy scale $\mu=2\,\rm{GeV}$, analogous pole contribution $(47-70)\%$, analogous $D_{10}$ contribution $(2-6)\%$, we have to postpone the continuum threshold parameter to much larger value $\sqrt{s_0}=11.75\pm0.10\,\rm{GeV}$, then we obtain the Borel window $T^2=(6.6-7.6)\,\rm{GeV}^2$, and the central values of the predicted mass and pole residue
 $M_Y=11.20\,\rm{GeV}$ and $\lambda_Y=2.13\times 10^{-1}\,\rm{GeV}^{6}$. The predicted mass  $M_Y=11.20\,\rm{GeV}$ is much larger than the experimental data   $M_Y=10752.7\pm5.9\,{}^{+0.7}_{-1.1}\,\rm{MeV}$ from the Belle    collaboration \cite{Belle-Y10750}.  The modified energy scale formula can
 enhance the pole contribution remarkably and improve the convergent behavior of the operator product expansion remarkably.   On the other hand, if we choose the typical energy scale $\mu=3\,\rm{GeV}$, analogous calculations lead to a mass about $10.42\,\rm{GeV}$, which is smaller than the S-wave hidden-bottom tetraquark masses $10.61-10.65\,\rm{GeV}$ \cite{Wang-Hidden-Bottom} and should be abandoned.

\begin{figure}
 \centering
 \includegraphics[totalheight=6cm,width=9cm]{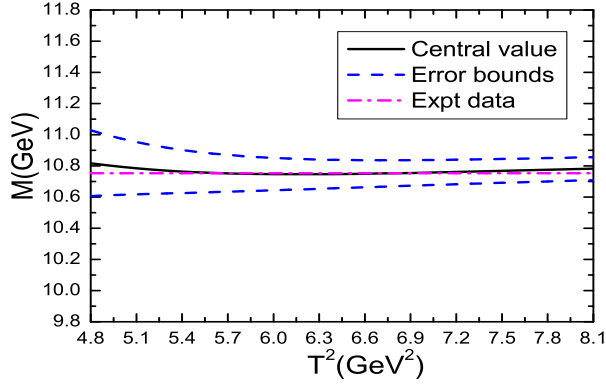}
 \caption{ The mass  of the vector hidden-bottom tetraquark candidate $Y(10750)$   with variation of the Borel parameter $T^2$.  }
\end{figure}

\begin{figure}
 \centering
 \includegraphics[totalheight=6cm,width=9cm]{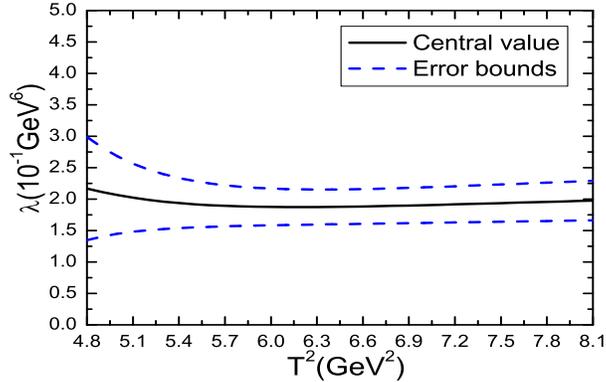}
 \caption{ The pole residue  of the  vector hidden-bottom tetraquark candidate $Y(10750)$   with variation of the Borel parameter $T^2$.  }
\end{figure}

\begin{figure}
 \centering
 \includegraphics[totalheight=6cm,width=9cm]{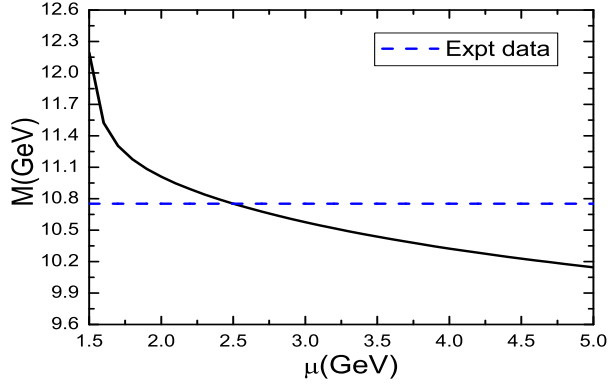}
 \caption{ The predicted mass  of the vector hidden-bottom tetraquark candidate $Y(10750)$  with variation of the energy scale $\mu$ for central values of the input parameters.  }\label{mass-mu}
\end{figure}

\section{The decay width of  the  vector tetraquark candidate $Y(10750)$}

Now we study the partial decay widths of the $Y(10750)$ as a vector hidden-bottom tetraquark candidate  with the three-point QCD sum rules, and write down the
  three-point correlation functions firstly,
\begin{eqnarray}
\Pi_{\nu}(p,q)&=&i^2\int d^4xd^4y e^{ipx}e^{iqy}\langle 0|T\Big\{J_{B}(x)J_{B}^{\dagger}(y)J_{\nu}^{\dagger}(0)\Big\}|0\rangle\, , \nonumber\\
\Pi^1_{\alpha\beta\nu}(p,q)&=&i^2\int d^4xd^4y e^{ipx}e^{iqy}\langle 0|T\Big\{J_{B^*,\alpha}(x)J^{\dagger}_{B^*,\beta}(y)J_{\nu}^{\dagger}(0)\Big\}|0\rangle\, , \nonumber\\
\Pi^1_{\mu\nu}(p,q)&=&i^2\int d^4xd^4y e^{ipx}e^{iqy}\langle 0|T\Big\{J_{B^*,\mu}(x)J_{B}^{\dagger}(y)J_{\nu}^{\dagger}(0)\Big\}|0\rangle\, , \nonumber\\
\Pi^2_{\mu\nu}(p,q)&=&i^2\int d^4xd^4y e^{ipx}e^{iqy}\langle 0|T\Big\{J_{\eta_b}(x)J_{\omega,\mu}(y)J_{\nu}^{\dagger}(0)\Big\}|0\rangle\, ,\nonumber\\
\Pi^3_{\mu\nu}(p,q)&=&i^2\int d^4xd^4y e^{ipx}e^{iqy}\langle 0|T\Big\{J_{\Upsilon,\mu}(x)J_{f_0}(y)J_{\nu}^{\dagger}(0)\Big\}|0\rangle\, ,\nonumber\\
\Pi^2_{\alpha\beta\nu}(p,q)&=&i^2\int d^4xd^4y e^{ipx}e^{iqy}\langle 0|T\Big\{J_{\Upsilon,\alpha}(x)J_{\omega,\beta}(y)J_{\nu}^{\dagger}(0)\Big\}|0\rangle\, ,
\end{eqnarray}
where
\begin{eqnarray}
J_{B}(x)&=&\bar{b}(x)i\gamma_5 u(x) \, , \nonumber\\
J_{B^*,\alpha}(x)&=&\bar{b}(x)\gamma_\alpha u(x) \, , \nonumber\\
J_{\eta_b}(x)&=&\bar{b}(x)i\gamma_5 b(x)\, ,\nonumber\\
J_{\omega,\mu}(y)&=&\frac{\bar{u}(y)\gamma_\mu  u(y)+\bar{d}(y)\gamma_\mu  d(y)}{\sqrt{2}}\, , \nonumber\\
J_{\Upsilon,\mu}(x)&=&\bar{b}(x)\gamma_\mu b(x)\, ,\nonumber\\
J_{f_0}(y)&=&\frac{\bar{u}(y)   u(y)+\bar{d}(y)   d(y)}{\sqrt{2}}\, ,\nonumber\\
J_\nu(0)&=&J_\nu^{(0,0)}(0)\, .
\end{eqnarray}

At the phenomenological side,  we insert  a complete set of intermediate hadronic states with the same quantum numbers as the current operators  into the three-point
correlation functions  and  isolate the ground state contributions  \cite{SVZ79,Reinders85},
\begin{eqnarray}\label{Coupling-first}
\Pi_{\nu}(p,q)&=& \frac{f_{B}^2m_{B}^4}{m_b^2}\frac{ \lambda_Y\,G_{YBB}}{\left(p^{\prime2}-m_Y^2\right)\left(p^2-m_{B}^2 \right)\left(q^2-m_{B}^2\right)}\,i\,\left(p-q\right)^\alpha\left(-g_{\alpha\nu}+\frac{p^{\prime}_{\alpha}p^\prime_{\nu}}{p^{\prime2}}\right)+\cdots\nonumber\\
&=&\Pi(p^{\prime2},p^2,q^2)\,\left( -ip_{\nu}\right)+\cdots \, ,
\end{eqnarray}

\begin{eqnarray}
\Pi^1_{\alpha\beta\nu}(p,q)&=& \frac{f^2_{B^*}m^2_{B^*}\,\lambda_Y\,G_{YB^*B^*}}{\left(p^{\prime2}-m_Y^2\right)\left(p^2-m_{B^*}^2 \right)\left(q^2-m_{B^*}^2\right)}(-i)\left(p-q\right)^\sigma \left(-g_{\nu\sigma}+\frac{p^{\prime}_{\nu}p^\prime_{\sigma}}{p^{\prime2}}\right) \left(-g_{\alpha\rho} +\frac{p_{\alpha}p_{\rho}}{p^2}\right)\nonumber\\
&&\left(-g_{\beta}{}^\rho+\frac{q_{\beta}q^{\rho}}{q^2}\right)+\cdots\nonumber\\
&=&\Pi(p^{\prime2},p^2,q^2)\,\left( i g_{\alpha\beta}p_\nu\right)+\cdots \, ,
\end{eqnarray}

\begin{eqnarray}
\Pi^1_{\mu\nu}(p,q)&=& \frac{f_{B}m_{B}^2}{m_b}\frac{f_{B^*}m_{B^*}\,\lambda_Y\,G_{YBB^*}\,\varepsilon_{\alpha\beta\rho\sigma}p^\alpha p^{\prime\rho}}{\left(p^{\prime2}-m_Y^2\right)\left(p^2-m_{B^*}^2 \right)\left(q^2-m_{B}^2\right)}\,i\,\left(-g_{\mu}{}^{\beta}+\frac{p_{\mu}p^{\beta}}{p^2}\right)\left(-g_{\nu}{}^{\sigma}+\frac{p^\prime_{\nu}p^{\prime\sigma}}{p^{\prime2}}\right)+\cdots\nonumber\\
&=&\Pi(p^{\prime2},p^2,q^2)\,\left(-i\varepsilon_{\mu\nu\alpha\beta}p^{\alpha}q^{\beta}\right)+\cdots \, ,
\end{eqnarray}

\begin{eqnarray}
\Pi^2_{\mu\nu}(p,q)&=& \frac{f_{\eta_b}m_{\eta_b}^2}{2m_b}\frac{f_{\omega}m_{\omega}\,\lambda_Y\,G_{Y\eta_b\omega}\,\varepsilon_{\alpha\beta\rho\sigma}q^\alpha p^{\prime\rho}}{\left(p^{\prime2}-m_Y^2\right)\left(p^2-m_{\eta_b}^2 \right)\left(q^2-m_{\omega}^2\right)}(-i)\left(-g_{\mu}{}^{\beta}+\frac{q_{\mu}q^{\beta}}{q^2}\right)\left(-g_\nu{}^\sigma+\frac{p^\prime_{\nu}p^{\prime\sigma}}{p^{\prime2}}\right)+\cdots\nonumber\\
&=&\Pi(p^{\prime2},p^2,q^2)\left(-i\varepsilon_{\mu\nu\alpha\beta}p^{\alpha}q^{\beta}\right)+\cdots \, ,
\end{eqnarray}

\begin{eqnarray}
\Pi^3_{\mu\nu}(p,q)&=& \frac{f_{\Upsilon}m_{\Upsilon}f_{f_0}m_{f_0}\,\lambda_Y\,G_{Y\Upsilon f_0}}{\left(p^{\prime2}-m_Y^2\right)\left(p^2-m_{\Upsilon}^2 \right)\left(q^2-m_{f_0}^2\right)}\,i\,\left(-g_{\mu\alpha}+\frac{p_{\mu}p_{\alpha}}{p^2}\right)\left(-g_\nu{}^\alpha+\frac{p^\prime_{\nu}p^{\prime\alpha}}{p^{\prime2}}\right)+\cdots\nonumber\\
&=&\Pi(p^{\prime2},p^2,q^2)\,\left(i g_{\mu\nu}\right)
+\cdots \, ,
\end{eqnarray}

\begin{eqnarray}\label{Coupling-final}
\Pi^2_{\alpha\beta\nu}(p,q)&=& \frac{f_{\Upsilon}m_{\Upsilon}f_{\omega}m_{\omega}\,\lambda_Y\,G_{Y\Upsilon\omega}}{\left(p^{\prime2}-m_Y^2\right)\left(p^2-m_{\Upsilon}^2 \right)\left(q^2-m_{\omega}^2\right)}(-i)\left(p-q\right)^\sigma \left(-g_{\nu\sigma}+\frac{p^{\prime}_{\nu}p^\prime_{\sigma}}{p^{\prime2}}\right) \left(-g_{\alpha\rho} +\frac{p_{\alpha}p_{\rho}}{p^2}\right)\nonumber\\
&&\left(-g_{\beta}{}^\rho+\frac{q_{\beta}q^{\rho}}{q^2}\right)+\cdots\nonumber\\
&=&\Pi(p^{\prime2},p^2,q^2)\,\left( i g_{\alpha\beta}p_\nu\right)+\cdots \, ,
\end{eqnarray}
where we have used the definitions for the decay constants and hadronic coupling constants,
\begin{eqnarray}
\langle 0|J_{B}(0)|B(p)\rangle&=&\frac{f_{B}m_{B}^2}{m_b}\, ,\nonumber\\
\langle 0|J_{B^*,\mu}(0)|B^*(p)\rangle&=&f_{B^*}m_{B^*}\xi^{B^*}_\mu\, ,\nonumber\\
\langle 0|J_{\eta_b}(0)|\eta_b(p)\rangle&=&\frac{f_{\eta_b}m_{\eta_b}^2}{2m_b}\, ,\nonumber\\
\langle 0|J_{\Upsilon,\mu}(0)|\Upsilon(p)\rangle&=&f_{\Upsilon}m_{\Upsilon}\xi^{\Upsilon}_\mu\, ,\nonumber\\
\langle 0|J_{\omega,\mu}(0)|\omega(p)\rangle&=&f_{\omega}m_{\omega}\xi^{\omega}_\mu\, ,\nonumber\\
\langle 0|J_{f_0}(0)|f_0(p)\rangle&=&f_{f_0}m_{f_0}\, ,
\end{eqnarray}

\begin{eqnarray}\label{CouplingC}
\langle B(p)B(q)|X(p^\prime)\rangle &=& -(p-q)^\alpha\xi_\alpha^{Y}\, G_{YBB}\, ,\nonumber\\
\langle B^*(p)B^*(q)|X(p^\prime)\rangle &=& (p-q)^\alpha\xi_\alpha^{Y}\xi^{B^* *}_\beta\xi^{B^* *\beta}\, G_{YB^* B^*}\, ,\nonumber\\
\langle B^*(p) B(q)|X(p^\prime)\rangle &=& -\varepsilon^{\alpha\beta\rho\sigma}\,p_\alpha \xi^{B^* *}_{\beta}p^{\prime}_\rho\xi_\sigma^{Y}\, G_{YBB^* }\, , \nonumber\\
\langle \eta_b(p)\omega(q)|X(p^\prime)\rangle &=& \varepsilon^{\alpha\beta\rho\sigma}\,q_\alpha \xi^{\omega*}_{\beta}p^{\prime}_\rho\xi_\sigma^{Y}\, G_{Y\eta_b \omega}\, ,\nonumber\\
\langle \Upsilon(p)f_0(q)|X(p^\prime)\rangle &=& -\xi^{*\alpha}_{\Upsilon}\xi_\alpha^{Y}\, G_{Y\Upsilon f_0}\, ,\nonumber\\
\langle \Upsilon(p)\omega(q)|X(p^\prime)\rangle &=& (p-q)^\alpha\xi_\alpha^{Y}\xi^{\Upsilon *}_\beta\xi^{\omega *\beta}\, G_{Y\Upsilon \omega}\, ,
\end{eqnarray}
 the   $\xi^{B^*}_\mu$, $\xi^{\Upsilon}_\mu$, $\xi^{\omega}_\mu$ and $\xi^{Y}_\mu$ are the polarization vectors of the  conventional  mesons and tetraquark candidate $Y(10750)$, respectively,  the
 $G_{YBB}$, $G_{YB^*B^*}$, $G_{YBB^*}$, $G_{Y\eta_b \omega}$,
 $G_{Y\Upsilon f_0}$ and  $G_{Y\Upsilon\omega}$   are the hadronic coupling constants.
In calculations, we observe that the hadronic coupling constant $G_{Y\Upsilon\omega}$ is zero at the leading order approximation, and we will neglect the process  $Y(10750)\to \Upsilon\,\omega\to \Upsilon\pi^+\pi^-\pi^0$.

We usually assign  the lowest scalar  nonet  mesons $\{f_0/\sigma(500),a_0(980),\kappa_0(800),f_0(980) \}$ as the tetraquark
states,  and assign the higher  scalar nonet mesons
$\{f_0(1370),a_0(1450),K^*_0(1430),f_0(1500) \}$ as the
conventional ${}^3P_0$ quark-antiquark  states \cite{Close2002,ReviewJaffe,ReviewAmsler2}. In this article, we assume $f_0=f_0(1370)$ with the symbolic quark structure $f_0(1370)=\frac{\bar{u}u+\bar{d}d}{\sqrt{2}}$.

We study the components  $\Pi(p^{\prime2},p^2,q^2)$ of the correlation functions in Eq.\eqref{Coupling-first}-\eqref{Coupling-final}, and  carry out the operator product expansion up to the vacuum condensates of dimension 5. We calculate both the connected and disconnected Feynman diagrams, take into account the perturbative terms, quark condensate and mixed condensate, and neglect the tiny contributions of the gluon condensate. Then we obtain the QCD spectral densities through dispersion relation,  match the hadron side with the QCD side of the components  $\Pi(p^{\prime2},p^2,q^2)$, perform double Borel transform with respect to $P^2=-p^2$ and $Q^2=-q^2$ by setting $p^{\prime2}=p^2$ in the hidden-bottom channels and $p^{\prime2}=4p^2$ in the open-bottom channels    to obtain the QCD sum rules for the hadronic coupling constants,
\begin{eqnarray}
&& \frac{f_{B}^2m_{B}^4}{m_b^2}\frac{ \lambda_Y\,G_{YBB}}{4\left(\widetilde{m}_Y^2-m_{B}^2\right)}\left[\exp\left(-\frac{m_{B}^2}{T_1^2}\right) -\exp\left(-\frac{\widetilde{m}_Y^2}{T_1^2}\right) \right]\exp\left( -\frac{m_{B}^2}{T_2^2}\right)\nonumber\\
 &&+\left(C_{Y^{\prime}B^+}+C_{Y^{\prime}B^-} \right)\exp\left(-\frac{m_{B}^2}{T_1^2} -\frac{m_{B}^2}{T_2^2}\right)\nonumber\\
&=&-\frac{1}{512\pi^4}\int_{m_b^2}^{s^0_{B}} ds \int_{m_b^2}^{s^0_{B}} du
\left(1-\frac{m_b^2}{s}\right)^2\left(1-\frac{m_b^2}{u}\right)^2\frac{m_b^2}{s}
\left(3s^2-5su-14s m_b^2+4u m_b^2\right) \nonumber\\
&& \exp\left(-\frac{s}{T_1^2}-\frac{u}{T_2^2}\right) \nonumber\\
&&-\frac{m_b\langle\bar{q}q\rangle}{192\pi^2} \int_{m_b^2}^{s^0_{B}} du\left(1-\frac{m_b^2}{u}\right)^2 \left(u+11m_b^2\right)
\exp\left(-\frac{m_b^2}{T_1^2}-\frac{u}{T_2^2}\right) \nonumber\\
&&+\frac{m_b\langle\bar{q}q\rangle}{192\pi^2}\int_{m_b^2}^{s^0_{B}} ds\left(1-\frac{m_b^2}{s}\right)^2 \left(3s-19m_b^2+\frac{4m_b^4}{s}\right)
\exp\left(-\frac{s}{T_1^2}-\frac{m_b^2}{T_2^2}\right) \nonumber\\
&&-\frac{m_b\langle\bar{q}g_{s}\sigma Gq\rangle}{768\pi^2}\int_{m_b^2}^{s^0_{B}} du\left(1-\frac{m_b^2}{u}\right)^2 \left(27-\frac{u+29m_b^2}{T_1^2}\right)
\exp\left(-\frac{m_b^2}{T_1^2}-\frac{u}{T_2^2}\right) \nonumber\\
&&-\frac{m_b^3\langle\bar{q}g_{s}\sigma Gq\rangle}{384\pi^2T_1^2}\int_{m_b^2}^{s^0_{B}} du\left(1-\frac{m_b^2}{u}\right)^2 \left(9-\frac{u+11m_b^2}{2T_1^2}\right)
\exp\left(-\frac{m_b^2}{T_1^2}-\frac{u}{T_2^2}\right) \nonumber\\
&&-\frac{m_b\langle\bar{q}g_{s}\sigma Gq\rangle}{768\pi^2}\int_{m_b^2}^{s^0_{B}} ds\left(1-\frac{m_b^2}{s}\right)^2 \left(1-\frac{4m_b^2}{s}\right)
\left(1+\frac{3s-m_b^2}{T_2^2}\right)\exp\left(-\frac{s}{T_1^2}-\frac{m_b^2}{T_2^2}\right) \nonumber\\
&&-\frac{m_b^3\langle\bar{q}g_{s}\sigma Gq\rangle}{384\pi^2T_2^2}\int_{m_b^2}^{s^0_{B}} ds\left(1-\frac{m_b^2}{s}\right)^2\left\{1-\frac{4m_b^2}{s}
-\frac{1}{2T_2^2}\left[18m_b^2-3s+m_b^2\left(1-\frac{4m_b^2}{s}\right)\right]\right\} \nonumber\\
&&\exp\left(-\frac{s}{T_1^2}-\frac{m_b^2}{T_2^2}\right) \nonumber\\
&&-\frac{m_b\langle\bar{q}g_{s}\sigma Gq\rangle}{768\pi^2}\int_{m_b^2}^{s^0_{B}} du\left(1-\frac{m_b^2}{u}\right)^2 \left(-9+\frac{7u+11m_b^2}{T_1^2}\right)
\exp\left(-\frac{m_b^2}{T_1^2}-\frac{u}{T_2^2}\right) \nonumber\\
&&-\frac{m_b\langle\bar{q}g_{s}\sigma Gq\rangle}{768\pi^2}\int_{m_b^2}^{s^0_{B}} ds\left(1-\frac{m_b^2}{s}\right)^2
\left\{\frac{4m_b^2}{s}-1+\frac{1}{T_2^2}\left[24m_b^2-3s+\left(1-\frac{4m_b^2}{s}\right) m_b^2 \right]\right\} \nonumber\\
&&\exp\left(-\frac{s}{T_1^2}-\frac{m_b^2}{T_2^2}\right) \nonumber\\
&&+\frac{m_b\langle\bar{q}g_{s}\sigma Gq\rangle}{256\pi^2}\int_{m_b^2}^{s^0_{B}} du\left(1-\frac{m_b^2}{u}\right) \left(5-\frac{4m_b^2}{u}\right)
\exp\left(-\frac{m_b^2}{T_1^2}-\frac{u}{T_2^2}\right) \nonumber\\
&&+\frac{m_b\langle\bar{q}g_{s}\sigma Gq\rangle}{128\pi^2}\int_{m_b^2}^{s^0_{B}} ds\left(1-\frac{m_b^2}{s}\right) \left(2-\frac{5m_b^2}{s}+\frac{2m_b^4}{s^2}\right)
\exp\left(-\frac{s}{T_1^2}-\frac{m_b^2}{T_2^2}\right) \nonumber\\
&&-\frac{m_b\langle\bar{q}g_{s}\sigma Gq\rangle}{256\pi^2}\int_{m_b^2}^{s^0_{B}} ds\left(1-\frac{3m_b^2}{s}+\frac{8m_b^4}{s^2}-\frac{2m_b^6}{s^3}\right)
\exp\left(-\frac{s}{T_1^2}-\frac{m_b^2}{T_2^2}\right) \, ,
\end{eqnarray}

\begin{eqnarray}
&&\frac{f_{B^*}^2 m_{B^*}^2 \lambda_Y G_{YB^* B^*}}{4\left(\widetilde{m}_Y^2-m_{B^*}^2\right)}
\left[\exp\left(-\frac{m_{B^*}^2}{T_1^2}\right)-\exp\left(-\frac{\widetilde{m}_Y^2}{T_1^2}\right)\right]
\exp\left(-\frac{m_{B^*}^2}{T_2^2}\right)  \nonumber\\
&&+C_{Y^{\prime}B^{*+}+Y^{\prime}B^{*-}}\exp\left(-\frac{m_{B^*}^2}{T_1^2}\right)\exp\left(-\frac{m_{B^*}^2}{T_2^2}\right) \nonumber\\
&=&\frac{1}{1536\pi^4}\int_{m_b^2}^{s^0_{B^*}} ds \int_{m_b^2}^{s^0_{B^*}} du
\left(1-\frac{m_b^2}{s}\right)^2 \left(1-\frac{m_b^2}{u}\right)^2 \frac{m_b^2}{s}
\left(2m_b^4+46s m_b^2-8u m_b^2-9s^2+5su\right) \nonumber\\
&&\exp\left(-\frac{s}{T_1^2}-\frac{u}{T_2^2}\right) \nonumber\\
&&+\frac{m_b\langle\bar{q}q\rangle}{192\pi^2}\int_{m_b^2}^{s^0_{B^*}} du\left(1-\frac{m_b^2}{u}\right)^2 \left(u-13m_b^2\right)
\exp\left(-\frac{m_b^2}{T_1^2}-\frac{u}{T_2^2}\right) \nonumber\\
&&+\frac{m_b\langle\bar{q}q\rangle}{192\pi^2}\int_{m_b^2}^{s^0_{B^*}} ds\left(1-\frac{m_b^2}{s}\right)^2 \left(3s-17m_b^2+\frac{2m_b^4}{s}\right)
\exp\left(-\frac{s}{T_1^2}-\frac{m_b^2}{T_2^2}\right) \nonumber\\
&&-\frac{m_b\langle\bar{q}g_{s}\sigma Gq\rangle}{1152\pi^2}\int_{m_b^2}^{s^0_{B^*}} du\left(1-\frac{m_b^2}{u}\right)^2
\left(45+\frac{7u-55m_b^2}{T_1^2}\right)\exp\left(-\frac{m_b^2}{T_1^2}-\frac{u}{T_2^2}\right) \nonumber\\
&&+\frac{m_b^3\langle\bar{q}g_{s}\sigma Gq\rangle}{384\pi^2T_1^2}\int_{m_b^2}^{s^0_{B^*}} du
\left(1-\frac{m_b^2}{u}\right)^2\left(-9+\frac{-u+13m_b^2}{2T_1^2}\right)\exp\left(-\frac{m_b^2}{T_1^2}-\frac{u}{T_2^2}\right) \nonumber\\
&&-\frac{m_b\langle\bar{q}g_{s}\sigma Gq\rangle}{576\pi^2}\int_{m_b^2}^{s^0_{B^*}} ds
\left(1-\frac{m_b^2}{s}\right)^2\left(1-\frac{4m_b^2}{s}\right)\left(1+\frac{3s-m_b^2}{T_2^2}\right)
\exp\left(-\frac{s}{T_1^2}-\frac{m_b^2}{T_2^2}\right) \nonumber\\
&&-\frac{m_b^3\langle\bar{q}g_{s}\sigma Gq\rangle}{384\pi^2T_2^2}\int_{m_b^2}^{s^0_{B^*}} ds
\left(1-\frac{m_b^2}{s}\right)^2\left(1-\frac{4m_b^2}{s}\right)\exp\left(-\frac{s}{T_1^2}-\frac{m_b^2}{T_2^2}\right) \nonumber\\
&&+\frac{m_b^3\langle\bar{q}g_{s}\sigma Gq\rangle}{768\pi^2T_2^4}\int_{m_b^2}^{s^0_{B^*}} ds
\left(1-\frac{m_b^2}{s}\right)^2\left[-3s+16m_b^2+\frac{2m_b^4}{s}+\left(1-\frac{4m_b^2}{s}\right)m_b^2\right]
\exp\left(-\frac{s}{T_1^2}-\frac{m_b^2}{T_2^2}\right) \nonumber\\
&&+\frac{m_b\langle\bar{q}g_{s}\sigma Gq\rangle}{768\pi^2}\int_{m_b^2}^{s^0_{B^*}} ds\left(1-\frac{m_b^2}{s}\right)\left(13-\frac{35m_b^2}{s}+\frac{16m_b^4}{s^2}\right)
\exp\left(-\frac{s}{T_1^2}-\frac{m_b^2}{T_2^2}\right) \nonumber\\
&&+\frac{m_b\langle\bar{q}g_{s}\sigma Gq\rangle}{768\pi^2}\int_{m_b^2}^{s^0_{B^*}} du\left(1-\frac{m_b^2}{u}\right)
\left(5-\frac{32m_b^2}{u}\right)\exp\left(-\frac{m_b^2}{T_1^2}-\frac{u}{T_2^2}\right) \nonumber\\
&&+\frac{m_b\langle\bar{q}g_{s}\sigma Gq\rangle}{2304\pi^2}\int_{m_b^2}^{s^0_{B^*}} ds
\left(4-\frac{9m_b^2}{s}+\frac{27m_b^4}{s^2}-\frac{10m_b^6}{s^3}\right)\exp\left(-\frac{s}{T_1^2}-\frac{m_b^2}{T_2^2}\right)\, ,
\end{eqnarray}

\begin{eqnarray}
&&\frac{f_{B^*}f_B m_{B^*}m_B^2 \lambda_Y G_{YB^* B}}{4m_b \left(\widetilde{m}_Y^2-m_{B^*}^2\right)}
\left[\exp\left(-\frac{m_{B^*}^2}{T_1^2}\right)-\exp\left(-\frac{\widetilde{m}_Y^2}{T_1^2}\right)\right]
\exp\left(-\frac{m_B^2}{T_2^2}\right)  \nonumber\\
&&+C_{Y^{\prime}B^{*+}+Y^{\prime}B^-}\exp\left(-\frac{m_{B^*}^2}{T_1^2}\right)\exp\left(-\frac{m_B^2}{T_2^2}\right) \nonumber\\
&=&\frac{m_b}{256\pi^4}\int_{m_b^2}^{s^0_{B^*}} ds \int_{m_b^2}^{s^0_{B}} du
\left(1-\frac{m_b^2}{s}\right)^2\left(1-\frac{m_b^2}{u}\right)^2\left(s+u-2m_b^2\right) \exp\left(-\frac{s}{T_1^2}-\frac{u}{T_2^2}\right) \nonumber\\
&&-\frac{\langle\bar{q}q\rangle}{96\pi^2}\int_{m_b^2}^{s^0_{B}} du\left(1-\frac{m_b^2}{u}\right)^2 \left(u-m_b^2\right)
\exp\left(-\frac{m_b^2}{T_1^2}-\frac{u}{T_2^2}\right) \nonumber\\
&&-\frac{\langle\bar{q}q\rangle}{96\pi^2}\int_{m_b^2}^{s^0_{B^*}} ds\left(1-\frac{m_b^2}{s}\right)^2 \left(s-m_b^2\right)
\exp\left(-\frac{s}{T_1^2}-\frac{m_b^2}{T_2^2}\right) \nonumber\\
&&+\frac{\langle\bar{q}g_{s}\sigma Gq\rangle}{288\pi^2T_1^2}\left(1+\frac{3m_b^2}{4T^2_1}\right)\int_{m_b^2}^{s^0_{B}} du\left(1-\frac{m_b^2}{u}\right)^2 \left(u-m_b^2\right)\exp\left(-\frac{m_b^2}{T_1^2}-\frac{u}{T_2^2}\right) \nonumber\\
&&+\frac{m_b^2\langle\bar{q}g_{s}\sigma Gq\rangle}{384\pi^2T_2^4}\int_{m_b^2}^{s^0_{B^*}} ds
\left(1-\frac{m_b^2}{s}\right)^2\left(s-m_b^2\right)\exp\left(-\frac{s}{T_1^2}-\frac{m_b^2}{T_2^2}\right) \nonumber\\
&&+\frac{\langle\bar{q}g_{s}\sigma Gq\rangle}{384\pi^2}\int_{m_b^2}^{s^0_{B}} du
\left(1-\frac{m_b^2}{u}\right)\left(-1+\frac{2m_b^2}{u}\right)\exp\left(-\frac{m_b^2}{T_1^2}-\frac{u}{T_2^2}\right) \nonumber\\
&&+\frac{\langle\bar{q}g_{s}\sigma Gq\rangle}{384\pi^2}\int_{m_b^2}^{s^0_{B^*}} ds
\left(1-\frac{m_b^2}{s}\right)\left(2-\frac{m_b^2}{s}\right) \exp\left(-\frac{s}{T_1^2}-\frac{m_b^2}{T_2^2}\right) \, ,
\end{eqnarray}

\begin{eqnarray}
&&\frac{f_{\eta_b}f_{\omega} m_{\eta_b}^2 m_{\omega} \lambda_Y G_{Y\eta_b \omega}}{2m_b \left(m_Y^2-m_{\eta_b}^2\right)}
\left[\exp\left(-\frac{m_{\eta_b}^2}{T_1^2}\right)-\exp\left(-\frac{m_Y^2}{T_1^2}\right)\right]
\exp\left(-\frac{m_{\omega}^2}{T_2^2}\right)  \nonumber\\
&&+C_{Y^{\prime}\eta_b+Y^{\prime}\omega}\exp\left(-\frac{m_{\eta_b}^2}{T_1^2}\right)\exp\left(-\frac{m_{\omega}^2}{T_2^2}\right) \nonumber\\
&=&\frac{m_b}{64\sqrt{2}\pi^4}\int_{4m_b^2}^{s^0_{\eta_b}} ds \int_{0}^{s^0_{\omega}} du
\sqrt{1-\frac{4m_b^2}{s}}\, u\, \exp\left(-\frac{s}{T_1^2}-\frac{u}{T_2^2}\right) \nonumber\\
&&-\frac{\langle\bar{q}q\rangle}{24\sqrt{2}\pi^2}\int_{4m_b^2}^{s^0_{\eta_b}} ds\sqrt{1-\frac{4m_b^2}{s}}\left(s-4m_b^2\right)
\exp\left(-\frac{s}{T_1^2}\right) \nonumber\\
&&+\frac{\langle\bar{q}g_{s}\sigma Gq\rangle}{72\sqrt{2}\pi^2T_2^2}\int_{4m_b^2}^{s^0_{\eta_b}} ds \sqrt{1-\frac{4m_b^2}{s}}\left(s-4m_b^2\right)\exp\left(-\frac{s}{T_1^2}\right) \nonumber\\
&&-\frac{\langle\bar{q}g_{s}\sigma Gq\rangle}{96\sqrt{2}\pi^2}\int_{4m_b^2}^{s^0_{\eta_b}} ds
\sqrt{1-\frac{4m_b^2}{s}}\exp\left(-\frac{s}{T_1^2}\right) \, ,
\end{eqnarray}

\begin{eqnarray}
&&\frac{f_{\Upsilon}f_{f} m_{\Upsilon} m_{f} \lambda_Y G_{Y\Upsilon f_0}}{m_Y^2-m_{\Upsilon}^2}
\left[\exp\left(-\frac{m_{\Upsilon}^2}{T_1^2}\right)-\exp\left(-\frac{m_Y^2}{T_1^2}\right)\right]
\exp\left(-\frac{m_{f_0}^2}{T_2^2}\right)  \nonumber\\
&&+C_{Y^{\prime}\Upsilon+Y^{\prime}f_0}\exp\left(-\frac{m_{\Upsilon}^2}{T_1^2}\right)\exp\left(-\frac{m_{f_0}^2}{T_2^2}\right) \nonumber\\
&=&\frac{m_b}{128\sqrt{2}\pi^4}\int_{4m_b^2}^{s^0_{\Upsilon}} ds \int_{0}^{s^0_{f_0}} du
\sqrt{1-\frac{4m_b^2}{s}}\, u\,  \left(u+2s-8m_b^2\right)\exp\left(-\frac{s}{T_1^2}-\frac{u}{T_2^2}\right) \nonumber\\
&&+\frac{\langle\bar{q}g_{s}\sigma Gq\rangle}{48\sqrt{2}\pi^2}\int_{4m_b^2}^{s^0_{\Upsilon}} ds\sqrt{1-\frac{4m_b^2}{s}}
\left(s+2m_b^2\right)\exp\left(-\frac{s}{T_1^2}\right)\, ,
\end{eqnarray}
where $\widetilde{m}_Y^2=\frac{m_Y^2}{4}$, the unknown functions $C_{Y^{\prime}B^+}+C_{Y^{\prime}B^-}$,
$C_{Y^{\prime}B^{*+}+Y^{\prime}B^{*-}}$,
$C_{Y^{\prime}B^{*+}+Y^{\prime}B^-}$,
$C_{Y^{\prime}\eta_b+Y^{\prime}\omega}$ and
$C_{Y^{\prime}\Upsilon+Y^{\prime}f_0} $ parameterize the transitions between the ground states ($B$, $B^*$, $\Upsilon$, $\eta_b$, $\omega$, $f_0(1370)$) and excited states $Y^\prime$.  For the definitions of the  unknown functions and technical details in calculations, we can consult Ref.\cite{WangZhang-solid}.

The  input parameters at the hadron side  are chosen  as
$m_{\Upsilon}=9.4603\,\rm{GeV}$,
$m_{\eta_b}=9.3987\,\rm{GeV}$,
$m_{\omega}=0.78265\,\rm{GeV}$,
$m_{B^+}=5.27925\,\rm{GeV}$,
$m_{B^{*+}}=5.3247\,\rm{GeV}$,
$m_{\pi^+}=0.13957\,\rm{GeV}$,
$\sqrt{s^0_{B}}=5.8\,\rm{GeV}$,
$\sqrt{s^0_{B^*}}=5.8\,\rm{GeV}$,
$\sqrt{s^0_{\Upsilon}}=9.9\,\rm{GeV}$,
$\sqrt{s^0_{\eta_b}}=9.9\,\rm{GeV}$ \cite{PDG},
$\sqrt{s^0_{\omega}}=1.3\,\rm{GeV}$, $f_{\omega}=215\,\rm{MeV}$ \cite{Wang-Zc4200},
$m_{f_0}=1.35\,\rm{GeV}$, $f_{f_0}=546\,\rm{MeV}$,  $\sqrt{s^0_{f_0}}=1.8\,\rm{GeV}$ (This work),
$f_{B}=194\,\rm{MeV}$, $f_{B^*}=213\,\rm{MeV}$  \cite{Wang-NPA-2004,Wang-heavy-decay},
$f_{\Upsilon}=649\,\rm{MeV}$ \cite{Latt-Upsilon}, $f_{\eta_b}=667\,\rm{MeV}$ \cite{Latt-etab}.

\begin{figure}
 \centering
  \includegraphics[totalheight=5cm,width=7cm]{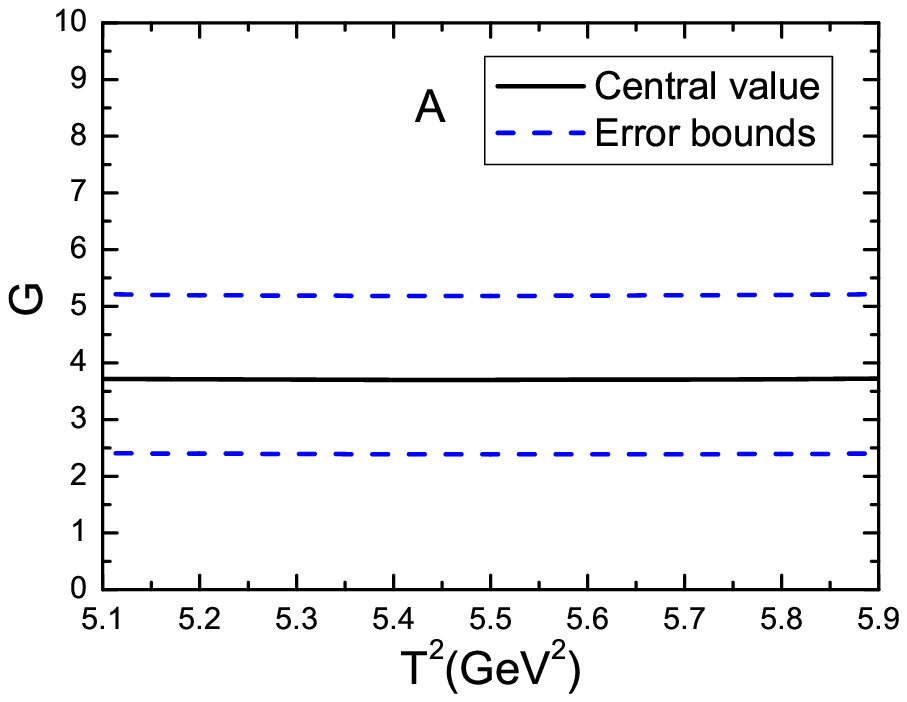}
  \includegraphics[totalheight=5cm,width=7cm]{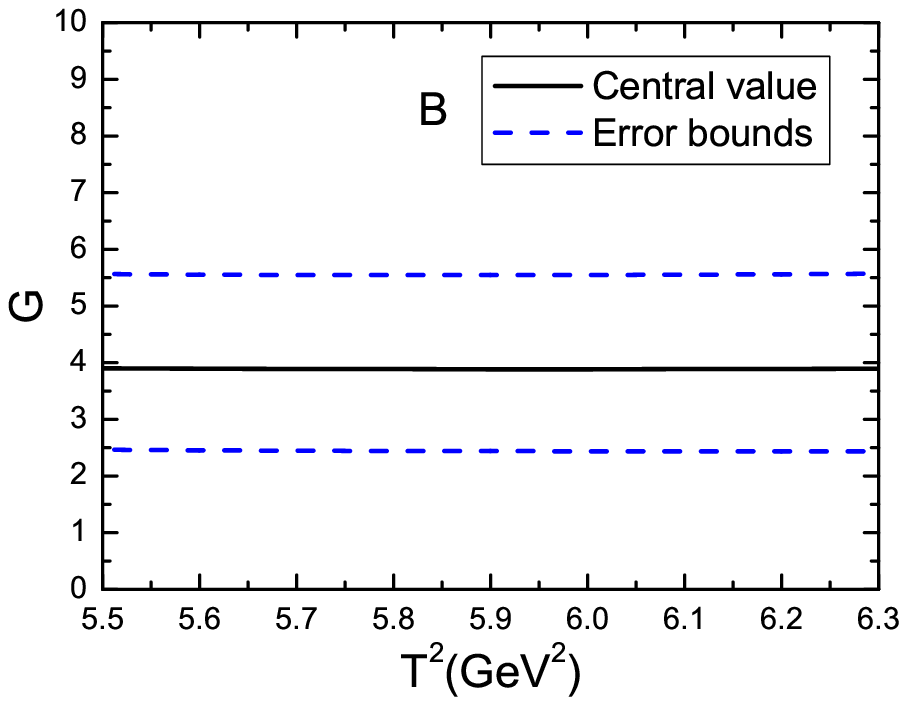}
  \includegraphics[totalheight=5cm,width=7cm]{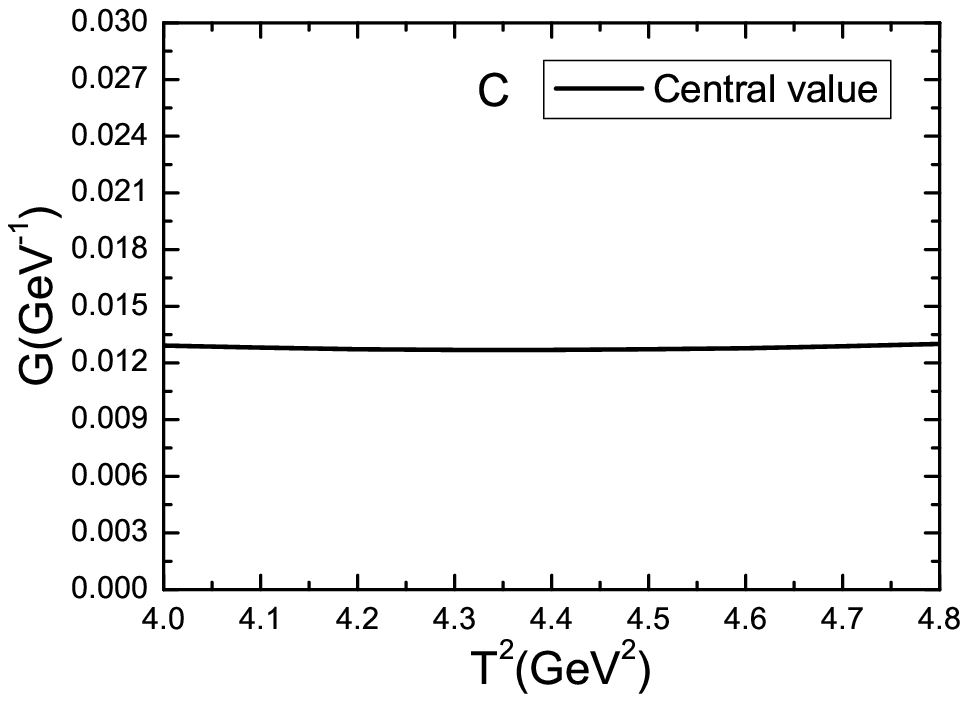}
  \includegraphics[totalheight=5cm,width=7cm]{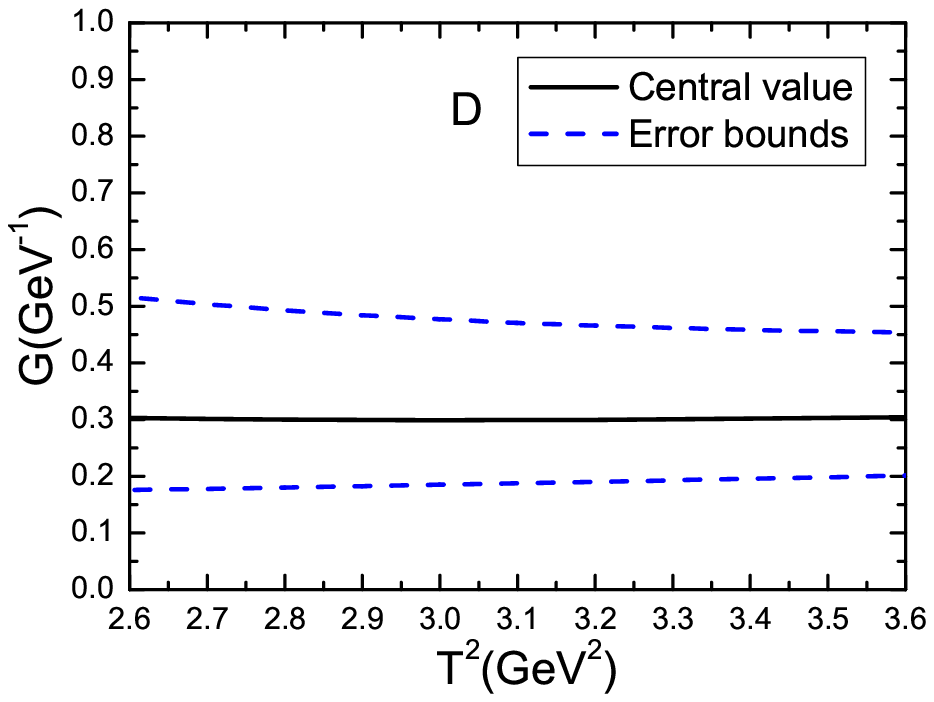}
  \includegraphics[totalheight=5cm,width=7cm]{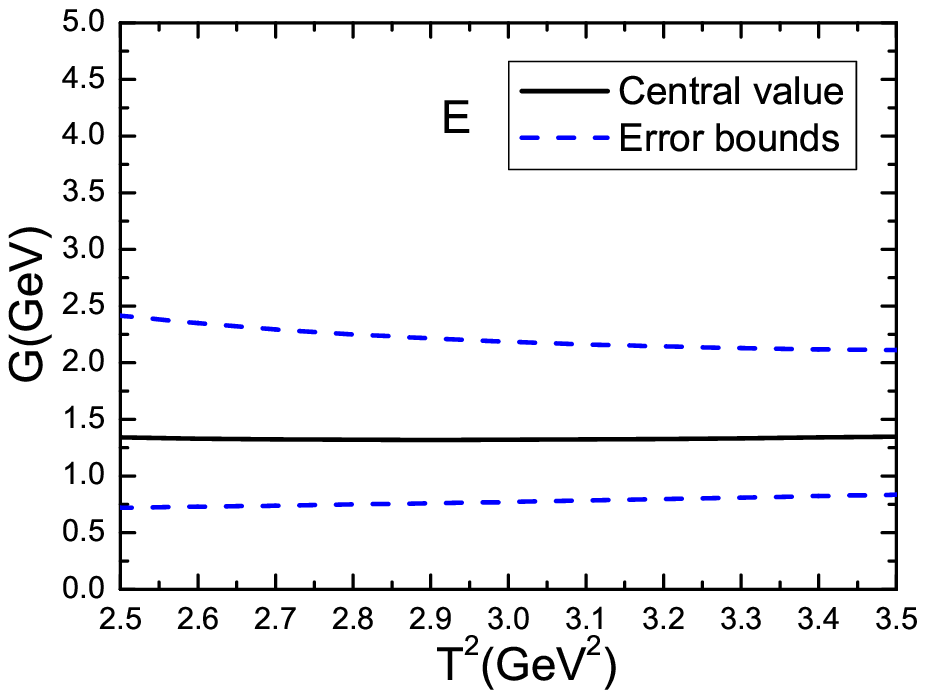}
     \caption{The hadronic coupling constants  with variations of the  Borel  parameters  $T^2$, the $A$, $B$, $C$, $D$ and $E$ denote the hadronic coupling constants $G_{YBB}$, $ G_{YB^*B^*}$,    $G_{YB^*B}$, $ G_{Y\eta_b\omega}$ and $G_{Y\Upsilon f_0}$, respectively.}\label{decay-constant}
\end{figure}

 We set the Borel parameters to be $T_1^2=T_2^2=T^2$ for simplicity in the QCD sum rules for the hadronic coupling constants  $G_{YBB}$, $G_{YB^*B^*}$, $G_{YB^*B}$ and
 $G_{Y\eta_b\omega}$, while in the QCD sum rules for the hadronic coupling constant $G_{Y\Upsilon f_0}$, the contribution in  the $u$ channel can be factorized out explicitly, we take the local limit, i.e. $T_2^2 \to \infty$ and $T_1^2=T^2$.
The unknown parameters are chosen as
$C_{Y^{\prime}B^+}+C_{Y^{\prime}B^-}=0.0441\,\rm{GeV}^8 $,
$C_{Y^{\prime}B^{*+}+Y^{\prime}B^{*-}}=0.0454\,\rm{GeV}^8 $,
$C_{Y^{\prime}B^{*+}+Y^{\prime}B^-}=0.00145  \,\rm{GeV}^7 $,
$C_{Y^{\prime}\eta_b+Y^{\prime}\omega}=0.0000125\,\rm{GeV}^7 $ and
$C_{Y^{\prime}\Upsilon+Y^{\prime}f_0}=0.00238\,\rm{GeV}^9 $
   to obtain  platforms in the Borel windows, which are shown in Table 1. In Fig.\ref{decay-constant}, we plot the hadronic coupling constants  $G$  with variations of the  Borel parameters $T^2$ in the Borel windows. The Borel windows $T_{max}^2-T^2_{min}=1.0\,\rm{GeV}^2$ for the hidden-bottom decays
   and  $T_{max}^2-T^2_{min}=0.8\,\rm{GeV}^2$ for the open-bottom decays, where the $T^2_{max}$ and $T^2_{min}$ denote the maximum and minimum of the Borel parameters.
We choose  the same intervals $T_{max}^2-T^2_{min}$ in all the QCD sum rules for the hadronic coupling constants in the two-body strong decays \cite{WangZhang-solid}, which work well for the decays of the $Z_c(3900)$, $X(4140)$, $Z_c(4600)$, $Y(4660)$, etc.

\begin{table}
\begin{center}
\begin{tabular}{|c|c|c|c|c|c|c|c|}\hline\hline
                                  &$T^2(\rm{GeV}^2)$   &$G$                                      &$\Gamma(\rm{MeV})$   \\ \hline

$Y(10750)\to B^+B^-$              &$5.1-5.9$           &$3.70^{+1.51}_{-1.31} $                  &$6.61^{+6.50}_{-3.85}$     \\ \hline

$Y(10750)\to B^{*+}B^{*-}$        &$5.5-6.3$           &$3.89^{+1.68}_{-1.45} $                  &$8.79^{+9.23}_{-5.33}$     \\ \hline

$Y(10750)\to B^{*+}B^{-}$         &$4.0-4.8$           &$\sim0.01\,\rm{GeV}^{-1}$                &$\sim 0.02$     \\ \hline

$Y(10750)\to \eta_b\, \omega$     &$2.6-3.6$           &$0.30^{+0.20}_{-0.12}\,\rm{GeV}^{-1}$    &$2.64^{+4.70}_{-1.69}$     \\ \hline

$Y(10750)\to \Upsilon f_0(1370)\to \Upsilon \pi^+\pi^-$  &$2.5-3.5$    &$1.32^{+1.10}_{-0.60}\,\rm{GeV}$     &$0.08^{+0.20}_{-0.06}$   \\ \hline

$Y(10750)\to \Upsilon \omega$    &                     &$\sim 0$                                 &$\sim 0$     \\ \hline\hline

\end{tabular}
\end{center}
\caption{ The Borel  windows, hadronic coupling constants and  partial decay widths of the $Y(10750)$ as the vector hidden-bottom tetraquark state. }
\end{table}

We take into account uncertainties of the input parameters, and obtain the  hadronic coupling constants, which are shown explicitly in Table 1 and Fig.\ref{decay-constant}.
Due to the tiny value of the  hadronic coupling constant $G_{YBB^*}$, we neglect the uncertainty.
Now we calculate the partial decay widths of the two-body strong decays $Y(10750)\to B^+B^-$, $ B^{*+}B^{*-}$,    $B^{*+}B^-$ and $ \eta_b\omega$ with formula,
\begin{eqnarray}
\Gamma\left(Y(10750)\to BC\right)&=& \frac{p(m_Y,m_B,m_C)}{24\pi m_Y^2} |T^2|\, ,
\end{eqnarray}
where $p(a,b,c)=\frac{\sqrt{[a^2-(b+c)^2][a^2-(b-c)^2]}}{2a}$, the $T$ are the scattering amplitudes defined in Eq.\eqref{CouplingC}, the numerical values of the
partial decay widths are shown in Table 1.

We assume  the three-body decays $Y(10750)\to \Upsilon f_0(1370)\to \Upsilon \pi^+\pi^-$  take place through a   intermediate virtual state $f_0(1370)$, and calculate the partial decay width,
\begin{eqnarray}
\Gamma(Y\to \Upsilon\pi^+\pi^-)&=&\int_{4m_\pi^2}^{(m_Y-m_{\Upsilon})^2}ds\,|T|^2\frac{p(m_Y,m_{\Upsilon},\sqrt{s})\,p(\sqrt{s},m_\pi,m_\pi)}{192\pi^3 m_Y^2 \sqrt{s} }\ , \nonumber\\
&=&0.08^{+0.20}_{-0.06}\,\rm{MeV}\, ,
\end{eqnarray}
where
\begin{eqnarray}
|T|^2&=&\frac{(M_Y^2-s)^2+2(5M_Y^2-s)M^2_{\Upsilon}+M^4_{\Upsilon}}{4M_Y^2 M_{\Upsilon}^2}G_{Y\Upsilon f_0}^2\frac{1}{(s-m_{f_0}^2)^2+s\Gamma_{f_0}^2(s)}G_{f_0\pi\pi}^2\, , \nonumber\\
\Gamma_{f_0}(s)&=&\Gamma_{f_0}(m_{f_0}^2) \frac{m_{f_0}^2}{s}\sqrt{\frac{s-4m_{\pi}^2}{m_{f_0}^2-4m_{\pi}^2}}\, , \nonumber\\
\Gamma_{f_0}(m_{f_0}^2)&=& \frac{G_{f_0\pi\pi}^2}{16\pi m_{f_0}^2}\sqrt{ m_{f_0}^2-4m_{\pi}^2 }\, ,
\end{eqnarray}
 $\Gamma_{f_0}(m_{f_0}^2)=200\,\rm{MeV}$ \cite{PDG}, the hadronic coupling constant $G_{f_0\pi\pi}$ is  defined by $\langle \pi^+(p)\pi^-(q)|f_0(p^\prime)\rangle=iG_{f_0\pi\pi}$. If we take the largest width $\Gamma_{f_0}(m_{f_0}^2)=500\,\rm{MeV}$ \cite{PDG}, we can obtain a slightly  larger
 partial decay width $\Gamma(Y\to \Upsilon\pi^+\pi^-)=0.11^{+0.27}_{-0.08}\,\rm{MeV}$.

Now it is easy to obtain the total decay width,
\begin{eqnarray}
\Gamma_{Y}&=& \Gamma\left(Y(10750)\to B^+B^-,\,B^0\bar{B}^0, \, B^{*+}B^{*-}, \, B^{*0}\bar{B}^{*0}, \, B^{*+}B^{-}, \, B^{+}B^{*-}, \right.\nonumber\\
 &&\left.B^{*0}\bar{B}^{0}, \, B^{0}\bar{B}^{*0},\, \eta_b\omega, \, \Upsilon\pi^+\pi^-\right)\, ,\nonumber\\
&=& 33.60^{+16.64}_{-9.45}\,{\rm{MeV}}\, ,
\end{eqnarray}
where we have assume the isospin limit for the $B$ and $B^*$ mesons.
	The predicted width $\Gamma_{Y}= 33.60^{+16.64}_{-9.45}\,{\rm{MeV}}$ is in excellent agreement with the experimental data $35.5^{+17.6}_{-11.3}\,{}^{+3.9}_{-3.3}\,\rm{MeV}$ from the Belle collaboration \cite{Belle-Y10750}, which also supports assigning the $Y(10750)$ to be the diquark-antidiquark type vector hidden-bottom tetraquark state.

In Ref.\cite{Zhong-Y10750}, Li et al assign the $Y(10750)$ and $\Upsilon(10860)$ to be the $5^3{\rm S}_1-4^3{\rm D}_1$ mixing states, the dominant components of the
$Y(10750)$ and $\Upsilon(10860)$ are the conventional bottomonium sates  $4^3{\rm D}_1$ and $5^3{\rm S}_1$, respectively. The decay mode $4^3{\rm D}_1 \to B^*B^*$ is the dominant mode, the decay mode $4^3{\rm D}_1 \to BB^*$ is sizable, while the decay mode $4^3{\rm D}_1 \to BB$ is nearly forbidden. In the present work, we assign the $Y(10750)$ to be the vector hidden-bottom tetraquark state, its dominant decay modes are $Y(10750)\to BB$ and $B^*B^*$, while the partial decay widths for the decays  $Y(10750) \to BB^*$ are tiny.
We can search for the $Y(10750)$ in the processes $Y(10750)\to B^+B^-$, $B^0\bar{B}^0$, $B^{*+}B^{*-}$,  $B^{*0}\bar{B}^{*0}$, $B^{*+}B^{-}$, $B^{+}B^{*-}$, $B^{*0}\bar{B}^{0}$,  $B^{0}\bar{B}^{*0}$, $\eta_b\omega$, $\Upsilon\pi^+\pi^-$ to diagnose the nature of the $Y(10750)$.

\section{Conclusion}
In this article, we take the scalar diquark and scalar antidiquark operators as the basic constituents, construct  the $C\gamma_5\otimes\stackrel{\leftrightarrow}{\partial}_\mu\otimes \gamma_5C$ type tetraquark current by introducing an explicit P-wave between the diquark and antidiquark constituents  to study  the $Y(10750)$ as a vector tetraquark state with the QCD sum rules. We carry out the operator product expansion  up to the vacuum condensates of dimension 10 in a consistent way, and use the modified  energy scale formula to choose the ideal energy scale of the QCD spectral density so as to extract the reasonable mass and pole residue. The  predicted mass $M_{Y}=10.75\pm0.10\,\rm{GeV}$ is in excellent agreement with (at least is compatible with) the experimental value  $M_Y=10752.7\pm5.9\,{}^{+0.7}_{-1.1}\,\rm{MeV}$ from the Belle    collaboration. Furthermore, we study the hadronic coupling constants in the two-body strong decays of the $Y(10750)$ with the three-point correlation functions by carrying out the operator product expansion up to the vacuum condensates of dimension $5$. We take into account both the connected and disconnected Feynman diagrams, and obtain the QCD sum rules for the hadronic coupling constants, then obtain the partial decay widths and total width. The predicted width  $\Gamma_{Y}= 33.60^{+16.64}_{-9.45}\,{\rm{MeV}}$ is in excellent agreement with the experimental data $35.5^{+17.6}_{-11.3}\,{}^{+3.9}_{-3.3}\,\rm{MeV}$ from the Belle collaboration.
 The present calculations favor  assigning the $Y(10750)$  as  the diquark-antidiquark type vector hidden-bottom tetraquark state with $J^{PC}=1^{--}$, which has a relative P-wave between the diquark and antidiquark constituents.

\section*{Acknowledgements}
This  work is supported by National Natural Science Foundation, Grant Number  11775079.

\end{document}